\documentclass[10pt]{iopart}
\input epsf

\begin{document}
\title[Simulation for plasma spectroscopy]{Molecular dynamics simulation for modeling plasma spectroscopy}
\author{B Talin\dag \footnote[5]{To whom correspondence should be addressed (btalin@piima1.univ-mrs.fr)},
E Dufour\dag, A Calisti\dag, M A Gigosos\ddag, M A
Gonz\'{a}lez\ddag, T del R\'{i}o Gaztelurrutia\S, and J W Dufty+}
\address{\dag\ Universit\'{e} de Provence, CNRS UMR 6633,
Centre Saint J\'{e}r\^{o}me, 13397 Marseille Cedex 20, France}
\address{\ddag\ Departamento de \'{O}ptica y F\'{i}sica Aplicada,
Facultad de Ciencias, Universidad de Valladolid, 47071 Valladolid,
Spain}
\address{\S\ Fisika Aplikatua Saila, Industri eta Telekomunikazio
Ingeniarien Goi Eskola Teknikoa, Urkijo Zumarkalea z{/}g, S-48013 Bilbo Spain}
\address{+\ Department of Physics, University of Florida, Gainesville, FL 32611, USA}

\begin{abstract}
The ion-electron coupling properties for a ion impurity in an electron gas
and for a two component plasma are carried out on the basis of a regularized
electron-ion potential removing the short-range Coulomb divergence. This
work is largely motivated by the study of radiator dipole relaxation in
plasmas which makes a real link between models and experiments. Current
radiative property models for plasmas include single electron collisions
neglecting charge-charge correlations within the classical quasi-particle
approach commonly used in this field. The dipole relaxation simulation based
on electron-ion molecular dynamics proposed here will provide means to
benchmark and improve model developments. Benefiting from a detailed study
of a single ion imbedded in an electron plasma, the challenging
two-component ion-electron molecular dynamics simulations are proven
accurate. They open new possibilities to obtain reference lineshape data.
\end{abstract}
\pacs{52.65.Yy, 32.70.Jz, 52.27.Aj}
\maketitle
\section{Introduction}
In this paper classical ion-electron plasma models are used to
carry out studies preliminary to line spectra simulations for ion
emitters. A classical plasma model involves point charges,
sometimes in a uniform charge background, interacting through
given additive binary forces and parameters characterizing
densities and temperatures. It is assumed that quantum mechanics
are either negligible or can be incorporated through appropriate
modification of the classical Coulomb potential. An example of
such a classical plasma is the one component plasma (OCP)
developed to study the statistical properties of non degenerate
ion fluid. The present study deals with ions of charge $Ze$ and
electrons. The inclusion of electrons requires essential quantum
mechanical effects to eliminate the divergence of the attractive
ion-electron Coulomb potential in the classical plasma. This is
accomplished by using a regularized Coulomb potential which
remains finite at short distances, representing quantum
diffraction effects \cite{Minoo}, and allowing the derivation of
all the desired properties using the laws of classical mechanics.
A detailed study of the time independent statistical properties of
an electron gas with an ion impurity has been carried out recently
on the basis of such a regularized electron-ion potential \cite
{Talin}. Two standard classical methods have been used for this
study: molecular dynamics (MD) and hypernetted chain approximation
(HNC) allowing to perform both cross comparisons and
interpretations of the results. A first interpretation of the
dynamical properties of the electron field autocorrelation
function at the impurity is reported in this issue
\cite{dufty-ilya}. Beside structure and correlation functions MD
can provide realistic representations of these stochastic electric
forces at the ion. Time dependent electric fields are the
necessary ingredients to describe Stark broadening (dipole
relaxation) of spectral lines emitted by ions in plasmas. It is of
great interest to describe these fields by simulation in order to
obtain lineshapes accounting for all the correlations between
charged particles. Such simulated lineshapes would provide
essential reference data to benchmark more efficient
phenomenological lineshape models developed for plasma diagnostics
via spectroscopy. Several recent models for line broadening have
been developed for investigations at new plasma conditions. They
suggest that part of the discrepancies found with experiments is
an inadequate description of the ion and electron perturber
coupling mechanisms. The main objective of our program is to
perform MD simulations with ions of charge $Z\geq 1$ and electrons
in order to derive all the relevant data required for lineshape
simulation accounting for all the interactions between charges.
The parameter range explored in this work is chosen to be
compatible with hot and dense plasmas diagnostics based on
spectroscopy. This is complementary to other MD simulation studies
of dense hydrogen \cite{zwick,knaup} with $Z=1$ and a much larger
electron-electron coupling strength. Two examples are provided to
show the potential for application of MD simulation to spectral
line broadening. In the next two sections the effects of
electron-electron correlations on electron broadening of spectral
lines is studied for a massive radiator in an electron gas. All
ion broadening is neglected in order to better isolate the
electron coupling effects. Next, in section 4 the more realistic
case of a radiator in a two component plasma (TCP) of electrons
and ions is considered. It is shown that separating the effects of
electron and ion broadening, as done in most current theories,
requires some care and interpretation.

\section{Ion impurity in an electron gas}
In this and the following section we consider a single ion impurity of
infinite mass imbedded in an electron gas with an additional uniform
positive background for charge neutrality. As mentioned above a more
detailed study has been reported elsewhere so only a few remarks will be
recalled here. The main statistical properties at the ion are derived using
both MD simulation and HNC approximation. The ion-electron Coulomb
regularized potential postulated is:
\begin{equation}
V_{ie}(r)=-Ze^{2}( 1-e^{-r/\delta }) e^{-r/\lambda}/r \label{2.1}
\end{equation}
where $\delta =(2\pi \hbar ^{2}/m_{e}k_{B}T)^{1/2}$ is the De
Broglie wavelength and $\lambda $ is a decay length long enough to
represent Coulomb correlations but short enough to allow periodic
boundary conditions (see below). Other parameters are the average
electron-electron distance $ r_{0}=(3/4\pi N_{e})^{1/3}$ defined
in terms of the electron density $N_{e}$ , the electron-electron
coupling constant $\Gamma $, the electron plasma frequency $\omega
_{p},$ and the ion-electron potential at the origin which, in
units of $r_{0}$, becomes: $\sigma =Z\Gamma /\delta $. For
simplicity, the OCP electron-electron potential is taken to be
$V_{ee}(r)=e^{2}e^{-r/ \lambda }/r$.\newline Molecular dynamics
simulations are carried out using a number $N$ of electrons in the
elementary cubic cell of size $c$ large enough to ensure the
following constraints: $Z<<N$, $\lambda >\lambda _{D}$ where
$\lambda _{D}$ is the Debye wavelength and $\lambda \leq c/2$.
These conditions are required in order that the Coulomb screening
mechanisms actually reduce the ion-ion interaction length.\newline
The hypernetted chain set of equations \cite{Rogers} is written
for the single ion case, i.e. for a vanishing ion density. As a
result, the ion-electron pair distribution function $g_{ie}(r)$ is
obtained by iteration as a function of the OCP direct
electron-electron correlation function $c_{ee}(r)$ calculated
separately. Both MD simulation and HNC approximation are
restricted to a limited range of values for the ion charge,
density, and temperature. Outside of this range electron trapping
and non-convergent iteration could occur invalidating the results.
For the parameter space considered MD
\begin{figure}
\begin{center}
\epsfbox{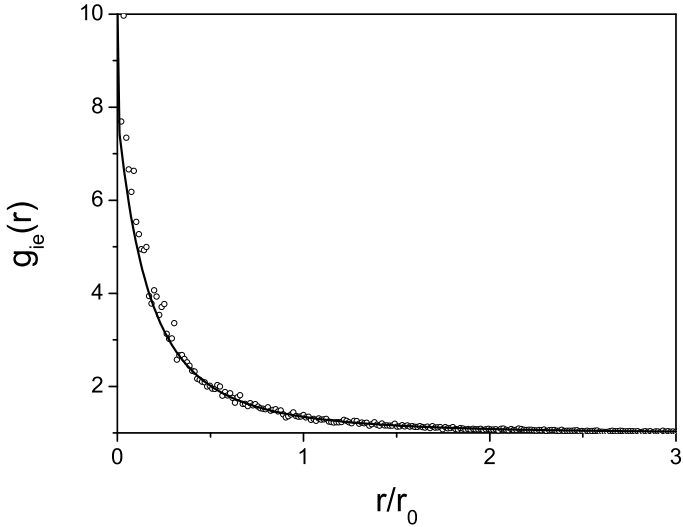}
\end{center}
\caption{\label{Figure 1} CVI impurity ion-electron pair
correlation function: $Z=5$, $N_{e}=10^{21}cm^{-3}$ and
$T_{e}=232000^{0}K$ from HNC: solid line, MD: circles}
\end{figure}
and HNC results are in good agreement as illustrated in Fig 1. The
electron accretion around the ion impurity shown by the shape of
the ion-electron pair distribution function is the expression of
screening mechanisms taking place due to the attractive forces. It
will be indirectly shown later that for these plasma parameters
the screening length is the Debye length.

\section{Implication for dipole relaxation mechanisms}
Most frequently models for Stark effect in plasmas are
semiclassical, i.e. they rely on the hypothesis that line
broadening can be given by the evolution of a quantum system - the
emitter - perturbed by the classical plasma electric micro-field.
In addition it is supposed that the emitter size is small in
respect to the micro-field space fluctuation so that this coupling
occurs via dipole interaction with the total electric micro-field
of the plasma. Due to the small electron-ion mass ratio, the Stark
effect resulting from slow ion and fast electron micro-fields are
generally considered separately. In this way the ion Stark effect
is approximated by a static field model, while electron broadening
is described as a dynamical process resulting from the high
frequency stochastic fluctuation of the electron micro-field. In
order to illustrate the implication of the present study for line
broadening attention is restricted in this section to electron
broadening alone, using the simple case of a massive emitter in an
electron gas. The usual approach for electron broadening for ion
emitters in plasmas relies on a binary collision model involving
independent electrons moving on straight trajectories at constant
velocity. Electron correlations are accounted for indirectly by
screening the electron-emitter potential at the Debye length. The
discussion of this approximation is not the purpose here, but
rather it is mentioned to motivate the following comparison. Two
MD simulations are performed to calculate the electron electric
field at the emitter, followed by calculation of the spectral line
shape: 1) the electrons are considered as free particles that do
not interact with each other but perturb the ion emitter with a
field obtained from the potential written in (\ref{2.1}) with
$\lambda =\lambda _{D}$; 2) the electrons have Coulomb
interactions with each other and couple to the emitter with a
field obtained using in (\ref{2.1}) with $\lambda =c/2$. The
latter is, of course, the correct treatment of correlations. Given
the fields from these MD calculations a lineshape simulation
follows. Although generally not coded in that way this simulation
process is equivalent, first, to solve numerically a stochastic
equation for the quantum system evolution operator:
\begin{eqnarray}
\frac{dU_{f_{i}}(t)}{dt}=-iL_{f_{i}}(t)U_{f_{i}}(t) \nonumber \\
\textbf{U}(t)=\{U_{f_{i}}(t)\}_{av}
=\frac{1}{N}\sum_{i=1}^{N}U_{f_{i}}(t) \label{3.1}
\end{eqnarray}
where $\{f_{1},f_{2}...f_{N}\}$ is a sample set of independent
perturbing field histories and $L_{f_{i}}(t)$ is an operator
involving the coupling of the emitter dipole with the external
field $f_{i}$. Then, the lineshape results from Fourier transform
of the emitter dipole autocorrelation function. It is written in
Liouville space as a scalar product in terms of the dipole vector
$|\textbf{d}\rangle\rangle$ and the equilibrium density matrix
$\rho$:

\begin{equation}
I(\omega)=\frac{1}{\pi}\int_{0}^{\infty}\langle\langle\textbf{d}|
\textbf{U}(t)|\rho\textbf{d}\rangle\rangle dt
\end{equation}
\begin{figure}
\begin{center}
\epsfbox{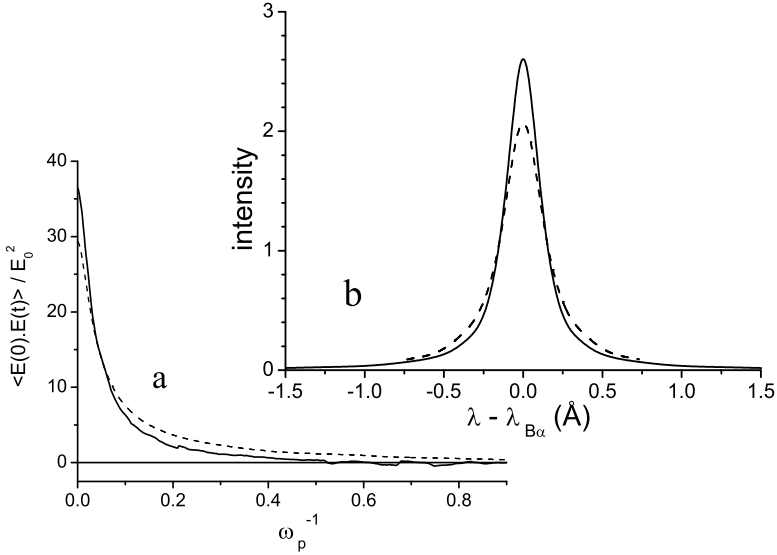}
\end{center}
\caption{\label{Figure 2} HeII impurity for independent electrons
(dashed line) and coupled electron (solid line): $Z=1$,
$N_{e}=10^{18}cm^{-3}$ and $T_{e}=60000K$. a) field
autocorrelation function, b) Balmer alpha line}
\end{figure}
These simulations require a fast integration process based on an
earlier work \cite{marco} carried out for hydrogenic lines. The
example reported is the Balmer alpha line emitted by an
hydrogen-like helium emitter in an electron gas with the density
$N_{e}=10^{18}cm^{-3}$ and the temperature $ T_{e}=60000K$. Figure
2 shows comparisons of independent electrons (case 1) and coupled
electrons (case 2). From the field autocorrelation functions
plotted on Fig 2-a $\langle \mathbf{E}(t)\cdot
\mathbf{E}(0)\rangle /E_{0}^{2}$, where brackets indicate a
classical Gibbs ensemble average and $E_{0}$ is the mean
electronic field, it can be inferred that the field de-correlation
rate of case 1 is lower than for case 2. An alternative
interpretation is that the field fluctuation rate is larger when
charge-charge correlations are accounted for than if they are not.
As the broadening mechanism is mainly a dynamical process, a
larger field fluctuation rate results in a narrower line, as shown
on Fig. 2b. However, this behavior found for an ionic charge $Z=1$
cannot be generalized straightforwardly since the broadening
effect depends not only on the field fluctuation rate but also on
the field covariance which is an increasing function of $Z$.

\section{Two component plasma MD simulations}
Intuitively, the electron microfield properties investigated above
for a single ion in an electron gas will be different if the
positive background assumed for charge neutrality is replaced by
mobile ions. In this section a charge-neutral two-component plasma
(TCP) of electrons and ions of charge $Ze$ will be considered.
Classical MD simulation is used to account for correlations among
all charges and cross comparisons performed for the single ion
case ensure the accuracy of these TCP simulations. The challenging
aspect of this study is to move slow and fast particles at the
same time. This requires simulations stable over a long period of
time, long enough to account for ion motion and based on a time
step compatible with electron motion. These simulations are
expensive as the particle number in the elementary cubic cell,
$N_{i}$ ions and $ZN_{i}$ electron, is large. However, a few
simulations have been carried out successfully. That for the same
plasma conditions than in Fig 1 is reported here. Before
proceeding it is worth contrasting such a MD simulation with other
methods of describing the TCP. Simulations involving ions and
electrons of a two component plasma have been carried out with the
so called quantum molecular dynamics \cite{Collins,dharma}. At
each time step the electrons are equilibrated about the ion
configuration (Born-Openheimer approximation) using an appropriate
density functional model. The ion-ion forces are then calculated
from the resulting electron density and these forces are used in
classical MD to move the ions. The process is repeated at each
time step. Alternatively, the electrons can be equilibrated by the
Carr-Parinello method. In both cases the electron dynamics is
fully equilibrated at each time step. The advantage of MD
simulation, although limited by the semi-classical potential, is
that both electrons and ions move dynamically according to the
given equations of motion. It appears that the two classes of
methods are complementary, with quantum molecular dynamics being
more appropriate for lower temperature and MD more appropriate at
higher temperatures.

\subsection{Screening effects}
The first point of interest is to verify that the expected ion-electron
screening mechanisms take place properly. As noted before, in the MD
elementary cubic cell the charge-charge interaction length is half the cell
size while the expected effective screening length (Debye) is only a
fraction of it. A comparison of the ion-ion pair distribution function for
the TCP and for the corresponding ion OCP is made. In the OCP system the
ion-ion Coulomb potential is screened at the Debye length, while such
screening must be generated by the electrons in the TCP simulations. Figure
3 shows a good agreement between the two simulations giving an indirect
evidence of the effectiveness of the ion-ion screening by the electrons.
Other calculations not reported here show that the agreement is preserved as
well when the temperature varies.
\begin{figure}
\begin{center}
\epsfbox{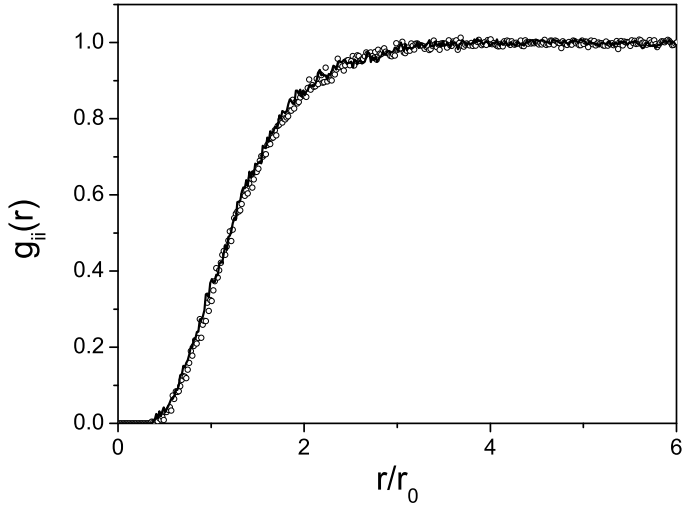}
\end{center}
\caption{\label{Figure 3}CVI ion-ion pair correlation
function, same conditions as in Fig 1; TCP: solid line, ion OCP:
circles}
\end{figure}

\subsection{Field dynamical properties}
Molecular dynamics simulation allows determination of statistical
data for each kind of particle involved. In the electron-ion TCP
two kinds of time dependencies coexist, the high frequency and the
low frequency dynamics related to electron motion and ion motion,
respectively. Both of them are of interest for probing or
discussing the simulation accuracy. For the high frequency data
the electron field autocorrelation function (FAF) at the ions has
been calculated. Figure 4-a shows a comparison of the field
autocorrelation function for a single ion in electrons and for the
TCP. It can be observed that for the TCP the FAF does not go to
$0$ as for the impurity case but, at the high frequency time scale
used for Fig 4-a, keeps a constant value that denotes the
occurrence of a low-frequency component in the electron field. At
a low-frequency dynamics time scale it will be seen below that, as
expected, this constant value goes to $0$. The low-frequency
component can be obtained averaging the electron field at the ions
on a period of time long enough for the field to loose correlation
of the high frequency fluctuation. Contrary to the single ion
case, a given ion is no longer a center of symmetry for the smooth
electron density that could result from the same high frequency
dynamics time average. Due to the screening mechanisms, the
structure of this smooth electron density is a blurred image of
the ion configuration which remains quasi-static at this time
scale. The primary consequence of the occurrence of a
low-frequency component in the TCP electron field is that it can
no longer be used in lineshape simulations. This will be a
motivation to carry out lineshape simulations accounting for the
total ion plus electron field. In order to investigate how the
low-frequency component is driven by the ion configuration several
FAF have been plotted on Fig 4-b: 1) the field autocorrelation
function of the low-frequency component, labelled "electron TCP";
2) the ion FAF for the ion OCP; 3) the ion FAF for the TCP. It can
be noted that the low-frequency component FAF goes to $0$ with
nearly the same rate than the ion FAF curves. For the TCP the
ion-ion interaction length is the cell size while for the OCP it
is the Debye length. This induces an inadequacy of the actual
ion-ion coupling and the pure ion-ion field. Therefore, the ion
TCP field appears useless when considered without the screening
electron field i.e. without the low-frequency component. This
explains the quite large discrepancy for the initial value of the
two ion FAF curves. Some differences occur also between the
dynamical behavior of the various partial fields. Further results
and analysis will be given elsewhere.
\begin{figure}
\begin{center}
\epsfbox{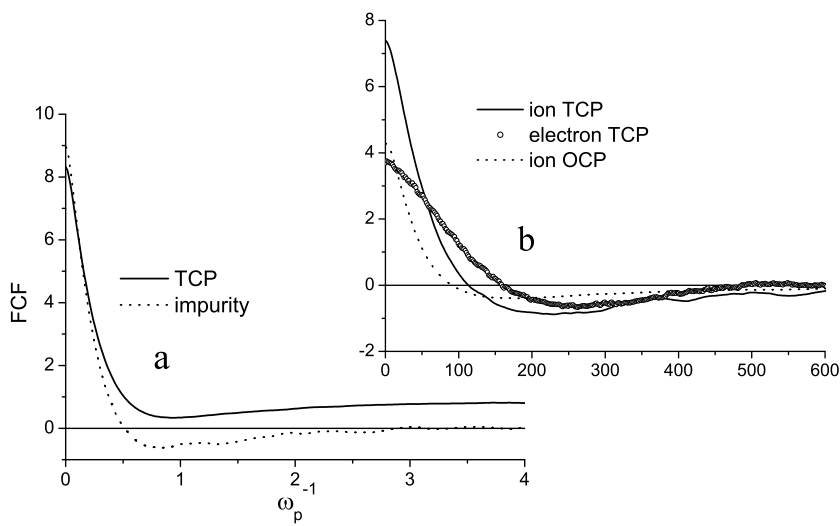}
\end{center}
\caption{\label{Figure 4} (same as in Fig 1 and 3) a) electron
FAF: single ion case and TCP, b) time averaged electron $\times5$,
ion OCP and ion TCP field auto correlation functions}
\end{figure}

\section{Discussion}
Classical molecular dynamics simulations of a two component
electron-ion plasma based on a regularized potential have been
carried out successfully. The step by step motion of slow ions and
fast electrons considered at the same time is inherently time
consuming. However, MD is a unique way for providing static and
time dependent statistical data accounting for all the
charge-charge couplings. Among them, the screening mechanisms of
forces take place accurately despite the limited size of the MD
simulation cell. The genuine MD microfields are directly
utilizable in semiclassical line-broadening calculations providing
the means to build a clear link between theoretical models and
plasma spectroscopy experiments. An example of such a calculation
has been reported here for the simple case of electron broadening
of a single ion in an electron gas. It shows the magnitude,
virtually observable, of the discrepancy induced by the
independent quasi-particle approximation. Line profile simulations
based on the two component plasma will be a unique way to perform
reference data for further developments of fast lineshape models.

\section{Acknowledgments}
Support for this research has been provided by the U. S.
Department of Energy grant No. DE-FG03-98DP0218. J. Dufty is
grateful for the support and hospitality of the University of
Provence.\\


\begin{thebibliography}{9}
\bibitem{Minoo} Minoo H, Gombert M and Deutsch C 1981 {\it Phys. Rev.} A {\bf 23} 2041
\bibitem{Talin} Talin B, Calisti A and Dufty J W 2002 {\it Phys.
Rev.} E {\bf 65}, 056406
\bibitem{dufty-ilya} Dufty J W, Pogorelov I, Calisti A and Talin B
this issue.
\bibitem{zwick} Pschiwul T and Zwicknagel G 2001 {\it Contrib. Plasma
Phys.} {\bf 41} 271
\bibitem{knaup} Knaup M, Zwicknagel G, Reihard P G, and Toepffer C
2000 {\it J. Phys. France IV}{\bf 10} 307
\bibitem{Rogers} Rogers F J 1984 {\it Phys. Rev.} A {\bf 29} 868
\bibitem{marco} Gigosos M A, Fraile J, and Torres F 1985 {\it Phys
Rev}A {\bf 31} 3509; Gigosos M A and Carde\~{n}oso v, 1987 {\it J
Phys} B {\bf 20} 6005; Gigosos M A and Carde\~{n}oso v, 1996 {\it
J Phys} B {\bf 29} 4795
\bibitem{Collins} Collins L, Kwon I, Kress J, Troullier N and
Lynch D 1995 {\it Phys. Rev.} E {\bf 52} 6202
\bibitem{dharma} Dharma-Wardana C in {\it Strongly Coupled Plasmas}, 275,  de
Witt H and Rogers F editors (NATO ASI Series, Plenum, NY, 1987).

\end{thebibliography}
\end{document}